# Growth and mortality of endangered land crabs (*Cardisoma guanhumi*) assessed through tagging with PITs and novel bootstrapped methods

RALF SCHWAMBORN[1*] & DENISE FABIANA DE MORAES-COSTA[2]

[1] *Oceanography Dept., Federal University of Pernambuco (UFPE), 50670-901 Recife, Brazil.*
   e-mail: rs@ufpe.br

[2] *Postgraduate Program in Animal Biology (PPGBA), Federal University of Pernambuco (UFPE), 50670-901 Recife, Brazil*

*Corresponding author

**Abstract.** The land crab *Cardisoma guanhumi* Latreille, 1828 is harvested in several countries in Latin America, and a critically endangered species. This is the first study to conduct bootstrapped tagging analysis (BTA) together with bootstrapped length-frequency analyses (BLFA). Crabs were sampled monthly in a mangrove patch at Itamaracá Island (Brazil), over 12 months, and marked with PIT tags. Both methods (BTA and BLFA) indicate very slow growth and $L_\infty$ far above $L_{max}$. BTA estimates were $K = 0.12$ $y^{-1}$ (95% CI: 0.024 to 0.26 $y^{-1}$), $L_\infty = 118$ mm (95% CI: 81 to 363 mm), $\Phi' = 1.23$ $\log_{10}(cm\ y^{-1})$ (95% CI: 0.86 to 1.36 $\log_{10}(cm\ y^{-1})$). Seasonality in growth was significant (p = 0.006, 95% CI for C: 0.15 to 0.93, median: C = 0.56). Pairs of K and $L\infty$ always followed narrow $\Phi'$ isopleths. Total mortality was $Z = 2.18$ $y^{-1}$ (95% CI = 1.7 to 4.5 $y^{-1}$). Slow growth and a very high Z/K ratio highlight the need for protective measures. BTA results were 2.2 to 3 times more precise than BLFA. Traditional length-based methods produced grossly biased results, indicating the urgent need for new, robust approaches and a critical reevaluation of long-standing methods and paradigms.

**Resumo**: **Crescimento e mortalidade de caranguejos terrestres (*Cardisoma guanhumi*) ameaçados de extinção, avaliados por marcação com PITs e novos métodos de bootstrap.** O caranguejo terrestre *Cardisoma guanhumi* Latreille, 1828 sofre capturas em vários países da América Latina, mesmo sendo uma espécie criticamente ameaçada. Este é o primeiro estudo a realizar análises de marcação e recaptura com bootstrap (BTA) junto com freqüências de comprimento com bootstrap (BLFA). Os caranguejos foram amostrados mensalmente em um manguezal na Ilha de Itamaracá (Brasil), durante 12 meses, e marcados com PITs. Observamos crescimento muito lento e $L_\infty$ muito acima de $L_{max}$. As estimativas do BTA foram: $K$ = 0,12 $y^{-1}$ (IC 95%: 0,024 a 0,26 $y^{-1}$), $L_\infty$ = 118 mm (IC 95%: 81 a 363 mm), $\Phi$ '= 1,23 $\log_{10}$ (cm $y^{-1}$) ( IC95%: 0,86 a 1,36 $\log_{10}$ (cm $y^{-1}$)). A sazonalidade no crescimento foi significativa. Os pares de K e L∞ sempre seguiram isolinhas Φ' estreitas. A mortalidade total foi de $Z$ = 2,18 a-1 (IC 95% = 1,7 a 4,5 $y^{-1}$). Crescimento lento e mortalidade alta indicam a necessidade de medidas de proteção. Os resultados do BTA foram 2,2 a 3 vezes mais precisos que o BLFA. Os métodos tradicionais baseados em comprimento produziram resultados com erros, indicando a necessidade urgente de abordagens novas e robustas e uma reavaliação crítica de métodos e paradigmas.

**Palavras-chave:** populações de crustáceos; dinâmica populacional; marcação e recaptura; fishboot; novos métodos

**Introduction**

Knowledge of growth and mortality is fundamental for the long-term management of populations and ecosystems. Length-frequency analysis (LFA) and tagging and are common methods to study populations of fish and invertebrates, but little has been done to assess and compare the inherent uncertainty of these methods.

Tagging is generally considered to be the most precise and accurate way to measure body growth in natural populations (Schmalenbach *et al.* 2011, Tang *et al.* 2014), since for each recaptured individual, its most recent size increment (dL/dt) is determined with great precision and accuracy. However, variability in growth will always lead to some uncertainty regarding the estimation of average growth parameters for the whole population (Hufnagl *et al.* 2012, Schwamborn *et al.* 2018b). This uncertainty may be substantial, since numbers of recaptured individuals are often very low, representing a minuscule fraction of the whole population. Tagging has several additional advantages and by-products, such as the determination of population size and the study of migrations and site fidelity (Moraes-Costa & Schwamborn 2018). However, tagging campaigns are often very costly and generally require huge efforts (Schmalenbach *et al.* 2011).

For data-poor populations, LFA can be a useful and cost-effective approach, since it requires only the regular capture and measurement of organisms (not necessarily purchase or sacrifice). This is probably the main reason why LFA remains hugely popular for fish stock assessments and population studies, since its first application in the late 19$^{th}$ century (Petersen, 1891). In most traditional methods for the analysis of data-poor situations (e.g., for most tropical invertebrates), only "one-estimate" results can be obtained for growth parameters, without any measures of uncertainty, which may rise questions whether traditional LFA is robust and replicable science at all (Schwamborn, 2018). More seriously, many common traditional length-based methods, such as the Powell-Wetherall plot (Wetherall 1986, Pauly 1986), may produce severely biased results (Hufnagl *et al.* 2012, Schwamborn, 2018), which indicates the need for new approaches and toolboxes.

It is well known that fitting the von Bertalanffy growth function (VBGF, von Bertalanffy 1934, 1938) to any given data set of length-frequencies is far from trivial (Shepherd *et al.* 1987), being considered an "elephantine" challenge for population ecology and stock assessment (Pauly & Greenberg 2013). The *a priori* definition of any fixed asymptotic size $L_\infty$ will drastically affect the results concerning the growth coefficient K and vice-versa (Shepherd *et al.* 1987, Kleiber & Pauly 1991, Pauly & Greenberg 2013, Schwamborn *et al.* 2018b). This makes the estimation of uncertainty in $L_\infty$ and K, each for itself, very doubtful, or even impossible. A simple and amply recommended solution to this fundamental issue in fitting a single VBGF to length data, is to use a fixed value of $L_\infty$, defined *a priori*, followed by the search for one optimum value of K (Sparre & Venema 1998, Mildenberger *et al.* 2017). This procedure can be assumed to be the "traditional length-based approach", as implemented in the widely used software packages FISAT II (Gayanilo *et al.* 1996, Gayanilo & Pauly 1997), ELEFAN in R (Pauly & Greenberg 2013), and *TropFishR* (Mildenberger *et al.* 2017). Yet, recent critiques of this traditional approach (Hufnagl *et al.* 2012, Schwamborn, 2018, Schwamborn *et al.* 2018b) have shown that it may lead to an overestimation of body growth rates and thus to an underestimation of the vulnerability to overfishing.

Here, we apply a new bootstrap-based approach, implemented in the *fishboot* software package (Schwamborn *et al.* 2018a), to a case study of "giant", or "blue" land crabs *Cardisoma guanhumi* (Latreille, 1825). These "giant" crabs occur from Florida to southern Brazil on the upper fringes of mangroves and adjacent lowlands, where they are generally the largest and most conspicuous brachyuran crab species in these coastal ecosystems (Tavares 2003), with $L_{max}$ reaching more than 100 mm carapace width in many parts of Florida and the Caribbean (Forsee & Albrecht 2012). They are considered an important food source in several countries, e.g., in Venezuela (Carmona-Suárez 2011), Brazil (Silva *et al.* 2014) and in Colombia (Cardona *et al.* 2019). Additionally to ongoing large-scale commercial capture by fishermen, these land crabs suffer a high predation pressure from numerous small mammals, such as crab-eating raccoons (Moraes-Costa & Schwamborn 2018). *C. guanhumi* is considered a critically endangered species in Brazil, the main threats being unsustainable harvesting and the destruction of mangroves and adjacent salt flats, e.g., for urbanization and shrimp farms.

Previous protection measures were the capture of males only, a minimum legal size (60 mm carapace width) and a seasonal closure from December to March. These measures, were, however, poorly respected and proved ineffective, leading to severe overharvesting. Thus, *C. guanhumi* has recently been listed as "critically endangered". This led to a complete prohibition of the capture of this species in Brazil, pushing thousands of crab harvesters into illegality, and leading to an enduring controversy among stakeholders and the scientific community. In spite of its huge socio-economic importance and its endangered status, little is known about growth and mortality of this species, except for three studies using traditional length-based approaches (Botelho *et al.* 2001, Silva *et al.* 2014, Moraes-Costa & Schwamborn 2016).

The objectives of this study were to assess growth and mortality of these endangered land crabs (*Cardisoma guanhumi*), including their inherent uncertainty, and to test and present a new bootstrapped approach for integrated length- and tagging-based analyses. We tested the hypothesis that the use of traditional methods (e.g., a *priori* fixing $L_\infty$) produces biased results, resulting in a underestimation of the vulnerability of this species to overharvesting.

**Materials and Methods**

**Study area**

The study area is a well-preserved mangrove patch located inside the National Center for Research and Conservation of Aquatic Mammals of the Brazilian ICMBio agency (CMA / ICMBio), at Itamaracá Island, Pernambuco State, Brazil (07 ° 48'36 "S, 034 ° 50'26"W at 07 ° 48'31"S, 034 ° 50'15"W).

The vegetation inside the study area is dominated by the mangroves *Rhizophora mangle* and *Conocarpus erectus*. Four sampling areas (A, B, C and D) were defined at the upper fringe of the mangrove, where *Cardisoma guanhumi* burrows were observed (Moraes-Costa & Schwamborn 2018). Additionally to the mangrove tree species *R. mangle* and *C. erectus,* typical beach vegetation also occurred at the sampling sites, such as *Terminalia catappa* and *Syzygium cumini*, forming a line of dense shrubs at the upper margin of the mangrove.

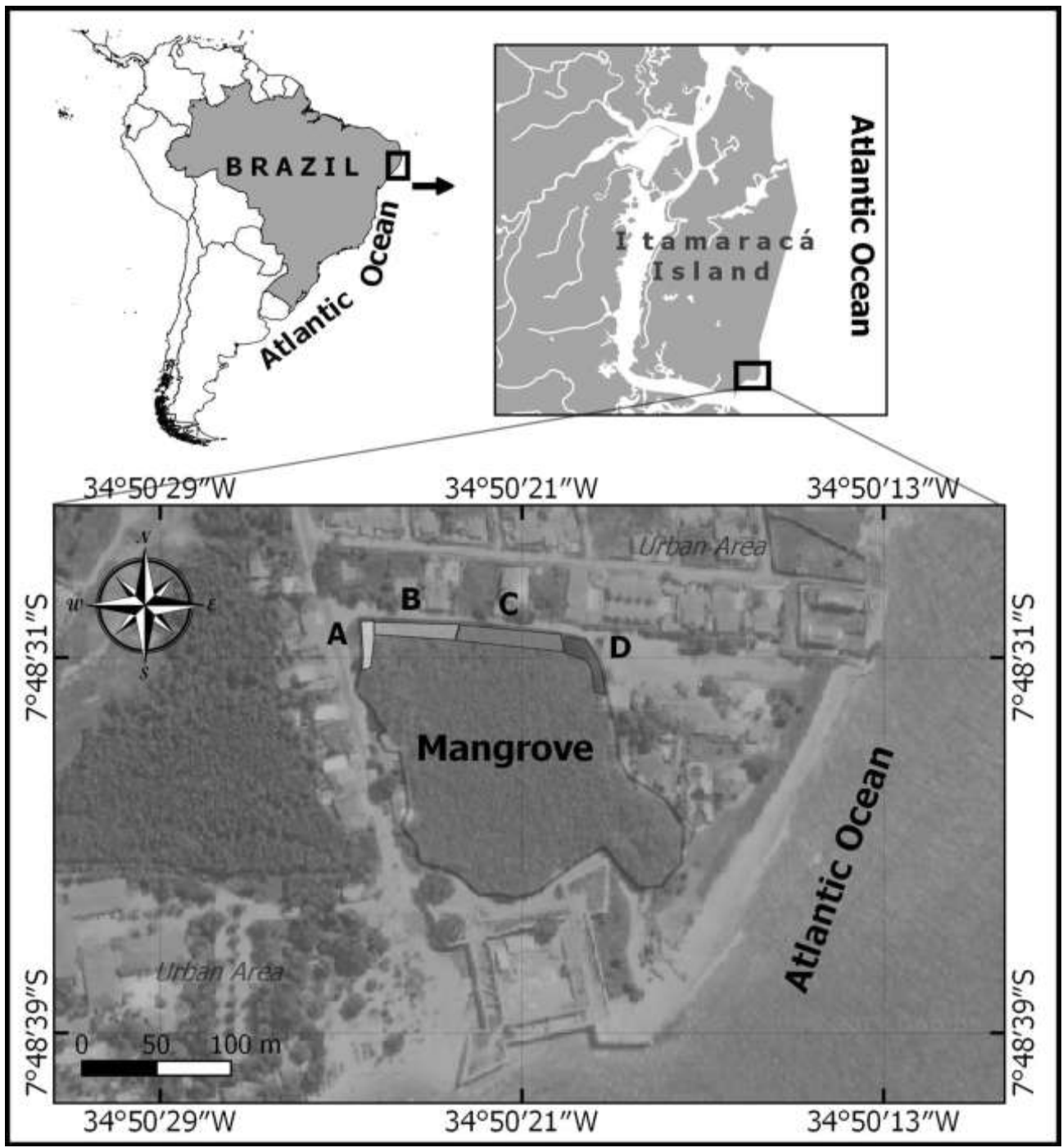

**Figure 1. Map of the study area identifying sampling sectors A, B, C and D, at the upper fringe of the CMA mangrove at Itamaracá Island, Pernambuco, State, Brazil**

According to the classification of Köppen (APAC 2016), the climate of Itamaracá Island, Pernambuco, is of the wet tropical type Ams' (Manso 2006). The rainy season in this region reaches from March to August, the peak dry season from November to January. Monthly accumulated rainfall values at Itamaracá during peak rainy season, in June and July 2015, were 234.1 mm and 291.0 mm, respectively. During the peak dry season months, rainfall in Itamaracá was zero in November 2015, 88.5 mm in December 2015 and 65.2 mm in January 2016 (APAC 2016). The locally measured differences in air temperature, between the highest monthly average (27.5 °C, January 2016) and the lowest monthly average (24.3 °C, July 2015), result in a seasonal temperature amplitude of 3.2° C.

**Sampling Strategy**

Monthly sampling was conducted during one year, from February 2015 to March 2016, along the upper fringe of the CMA mangrove. Cylindrical traps, identical to those used by artisanal fishermen, were built with plastic bottles and cans. Inside each trap, a pineapple fragment was used as bait, as done in regular artisanal harvesting. A total of 70 artisanal traps were built for this study. Each trap was positioned at the entrance of a burrow for forty-eight hours and evaluated every two hours. The individuals captured in each sector were distributed into twelve plastic boxes of 70 x 30 cm and allocated by sector and size group (maximum: 15 individuals per box, the bottom of the boxes covered with humid mangrove branches as to avoid stress and aggression), prior to measuring and tagging. All individuals were measured (carapace width, length and height) weighed, sexed, and released. Sex ratio (ratio of captured males: females) was tested for a significant difference from equality by a simple Chi-square test (Zar 1996).

**Tagging with PITs**

Of the 1,078 individuals captured, 291 individuals (153 males and 138 females) were tagged with passive integrated transponder tags (PIT, Nanotransponder tags, Trovan, model ID 100 A, dimensions: 1.25 mm x 7.0 mm). The tagged individuals had a carapace width of 24.4 to 59.5 mm (standard deviation: 7.4 mm).

PITs were always inserted into the ventral part of the carapace through the base of the fourth pereiopod, by injection with a specific syringe-type applicator. Each PIT has a unique numbering, which can only be obtained through a specific reader (Vantro Systems, model GR250).

To assess tag loss, a heat mark (quick branding with a soldering iron) was made on the upper part of the carapace (Diele & Koch 2010), which served as control at the time of recapture, indicating a tagged individual.

**Sex-specific growth and mortality**

Growth in size (carapace width, mm) was described using the von Bertalanffy growth function (VBGF, von Bertalanffy 1934, 1938), based on analyses of length-frequency distributions (LFDs) and on mark-recapture with PITs. The shape of the VBGF is mainly determined by two parameters: the growth constant K and the asymptotic length $L_\infty$. The growth performance index $\Phi'$ ("Phi-prime", Pauly & Munro 1984) was used to obtain a proxy that integrates $L_\infty$ and K, where $\Phi' = \log_{10}(K(y^{-1})) + 2 \log_{10}(L_\infty(cm))$.

Total mortality Z was estimated using the length-converted catch curve (LCCC) method (Baranov 1918, Ricker 1975, Pauly 1983, 1984a, 1984b). Differences between males and females regarding all relevant population parameters (mean size, median size, mean growth increments, K, $L_\infty$, Z, etc.) were tested at $p_{crit}$ = 0.05 using a non-parametric permutation test (function *independence_test* in the R package *coin*, Hothorn *et al.* 2006).

**Length-frequency analyses**

All individuals were grouped into size class intervals of 2 mm as to obtain monthly length-frequency distributions (LFDs). Growth parameters $L_\infty$ and K were estimated directly, based on the monthly LFD plots, using common traditional length-based methods, such as ELEFAN I (Eletronic LEngth-Frequency ANalysis, Pauly & David 1981) method, inserted in the FISAT II (Gayanilo *et al.* 1996, Gayanilo & Pauly 1997) and *TropFishR* (Mildenberger *et al.* 2017) software packages. To detect peaks (i.e., cohorts), ELEFAN I uses a moving average (MA) smoothing function, where the original LFDs (black bars in Fig. 1a) are transformed as differences between the smoothed curve and the original, resulting in a sequence of positive (peaks) and negative (troughs) values (black and white bars in Fig. 1b). The moving average span (MA = 7) for ELEFAN I was chosen based on the rule of thumb suggested by Taylor & Mildenberger (2017), where MA should be approximately equal to the number of bins spanning the youngest cohorts.

Additionally, the Bhattacharya (1967) method, also a part of the FISAT II package, was used to determine the peaks of the cohorts (green dots in Fig. 1a). This common method decomposes the frequency distributions into normally distributed cohorts with precise, unique peaks. A VBGF curve can be constructed by manually connecting these Bhattacharya peaks (Sparre & Venema 1998).

**The Powell-Wetherall method**

A simple approach is generally recommended within common traditional length-based methods, to fit a VBGF curve to LFD data, which consists in quickly determining $L_\infty$ and then conducting a detailed search for an optimum value of K, using a fixed value for $L_\infty$ (Gayanilo *et al.* 1995, 2005, Mildenberger 2017). The first step is to obtain a single estimate of $L_\infty$, either by using the Powell-Wetherall method (P-W method, Wetherall 1986, Pauly 1986, Schwamborn 2018), or directly based on the largest organism in the sample ("Lmax approach" or maximum-length approach, Mathews & Samuel 1990, Schmalenbach *et al.* 2011, Schwamborn 2018). Here, the P-W plot method, also called the "modified Wetherall method" (Wetherall 1986, Pauly 1986) was used as an additional and independent method to estimate $L_\infty$ and the mortality/growth (i.e., Z / K) ratio. The P-W method, originally proposed by Wetherall (1986), and modified by Pauly (1986), is based on partitioning a catch curve in consecutive cutoff lengths. The parameters (intercept and slope) of a linear regression are used to calculate the Z / K ratio and $L_∞$ (Pauly 1986). Different versions of the original and modified P-W methods were tested with the *C. guanhumi* LFD data, using FISAT II (Gayanilo *et al.* 1995, 2005), "ELEFAN in R" (Pauly & Greenberg 2013) and "TropFishR" (Mildenberger *et al.* 2017).

Two widely recommended (Sparre & Venema 1998) and commonly used, traditional LFA approaches were used for determining the growth coefficient K: ELEFAN I (Pauly & David 1981) with K-Scan (scanning a series of K values, using a fixed $L_∞$ value, previously obtained from the P-W method, Gayanilo *et al.* 1995, Moraes-Costa & Schwamborn 2016), and Response Surface Analysis (RSA, Gayanilo *et al.* 1995).

These methods were used by applying the widely used FISAT II software (Gayanilo *et al.* 2005) and, for comparison, the R package *TropFishR* (Mildenberger *et al.* 2017). In *TropFishR*, the genetic search algorithm ELEFAN_GA (Mildenberger *et al.* 2017) was used to fit growth models to the monthly LFDs, using precision-optimized search settings (e.g., maxiter = 100), and searching within a nearly unconstrained (i.e., extremely wide) search space (e.g., $L_∞$ from 0.5 * Lmax to 15 * Lmax, K from 0.01 to 1, and C from 0 to 1). Only one single curve fit attempt and one fixed seed value were used for ELEFAN_GA. Then, multiple, automatically repeated fit attempts, using the same, original LFDs, using only different seed values (i.e., conducting a partial bootstrap *sensu* Schwamborn *et al.* 2018b) were conducted using the function ELEFAN_GA_boot (Schwamborn *et al.* 2018a).

**Tagging-based analyses**

Body growth was also investigated by analyzing data obtained from mark-recapture. The input data were individual size increments, that is, two sets of paired variables: time lags (dt) and size differences (dL), for each recapture event. Increments in carapace width (dL/dt, in mm year$^{-1}$) were inserted into the *fishmethods* R package (Nelson 2018). The *grotag* function (Kienzle & Nelson 2018) was then applied to these increments, using the nonlinear adjustment method proposed by Francis (1988). The individual growth increments and the growth parameters K and $L_∞$ obtained with *grotag* were then used as inputs to plot these increments on a VBGF curve (Fig. 2) using the *growthTraject* function (Hoenig, 2018) within the *fishmethods* package (Nelson 2018). For comparison, we also tested the use of the Munro and Gulland Holt methods, using the *growth_tagging* function within the *TropFishR* package (Mildenberger *et al.* 2017).

**Seasonal growth**

Additionally to the common VBGF, the seasonally oscillating von Bertalanffy growth function was also used in this study (soVBGF, Ursin 1963a, 1963b, Pitcher & MacDonald 1973, Pauly & Gaschütz 1979, Francis 1988, Somers 1988). The soVBGF has two additional parameters: the seasonal amplitude "C" (Pauly & Gaschütz 1979) and a location parameter (e.g., winter point, WP). The seasonal amplitude "u" in Francis (1988) given by the function *grotag* in the R packge *fishmethods* is identical to the seasonal amplitude "C" in Pauly & Gaschütz (1979) and in ELEFAN I (Pauly & David, 1981). If $u = 0$, there is no seasonal variation, if $u = 1$, growth is zero in winter. The winter point "WP" of minimum growth in ELEFAN I is the opposite of "w" (date of maximum growth, *i.e.*, "summer point") estimated from tagging data using *grotag* (Francis 1988). WP was obtained from *grotag* outputs by WP = w + 0.5.

Two statistical procedures were used to verify whether there is a significant seasonality in growth. First, general additive models (GAM, Hastie & Tibshirani 1990) were applied to monthly growth time series obtained from mark-recapture. GAM models were built for raw growth increments (dl/dt) and for the residuals of the non-season VBG curve, using the *gam* package (Hastie 2018) in R at $p_{crit} = 0.05$.

Also, a Mann-Whitney U-test ($p_{crit} = 0.05$) was conducted to test for significant differences of the median increments (dL / dt) between rainy and dry season (Zar 1996). This test was based only on increments with less than 65 days of duration, that is, only increments with one and two months intervals were used for this seasonality analysis. Rainy season growth: increments (dL / dt) in carapace width (mm year$^{-1}$) of individuals recaptured from July to October 2015 (marked May to September 2015. Rainy season growth: increases (dL / dt) of carapace width (mm / year) of individuals recaptured in May 2015, and recaptured from December 2015 to March 2016 (marked April 2015 and marked October 2015 to February 2016).

**Integration of length-based and tagging-based methods into a seasonal growth curve**

Once a set of reliable estimates of K and $L_\infty$ had been obtained through tagging-based methods (*fishmethods* package), a new evaluation of the LFDs was made using a multi-step approach applying different ELEFAN I-based routines, to verify whether the growth curve obtained by tagging can be applied to the monthly LFDs, and to obtain final best fit estimates for growth parameters, by fine tuning and repeated fitting.

In this final step, growth, mortality (length-converted catch curve LCC, Pauly 1983, 1984a, 1984b) and recruitment (Pauly 1982) were assessed using FISAT II and *TropFishR* (Gayanilo *et al.* 2005, Mildenberger 2017). Recruitment patterns were also assessed by plotting the monthly abundance of small individuals (CW < 30 mm).

A new attempt was made to fit the VBGF growth curve to the monthly LFDs data, based on the K and $L_\infty$ estimates obtained with the *fishmethods* package, using ELEFAN I (applying the function *ELEFAN_GA* in the *TropFishR* package) to estimate only the SS (starting sample) and SL (starting length), with fixed K and $L_\infty$, as to obtain a single preliminary fit of the non-seasonal and seasonal VBGF curves to the monthly LFD data. Finally, the function *ELEFAN_GA_boot* in the *fishboot* package (Schwamborn *et al.* 2018a) was used to determine the final best fit parameters of the soVBGF curve (best fit values for K, $L_\infty$, C and ts) based on 1,000 repeated independent fitting attempts.

**Uncertainty in growth (BTA and BLFA)**

Confidence intervals (CIs) for growth parameters were obtained by bootstrapping two datasets: mark-recapture growth increments and monthly LFDs (Efron 1979, 1987, Schwamborn *et al.* 2018b). For bootstrapping tagging analysis (BTA), in each run, a random sample (of same size as the original data) was taken from the size increment data, with replacement. The *grotag* function (Kienzle & Nelson 2018) within the *fishmethods* package was then applied to these resampled increments, based on the nonlinear adjustment method proposed by Francis (1988), as to obtain unique values of K, $L_\infty$ and Φ' for each bootstrap run. Simple random sampling was repeatedly applied with replacement, using the standard *sample* function in R (at least 1,000 runs). This bootstrap routine for growth increments was built into the function *grotag_boot* within the new *fishboot* package (Schwamborn *et al.* 2018a). 95% CIs were then obtained from percentiles of the posterior distributions for K, $L_\infty$ and Φ'. Bivariate confidence contour envelopes (Fig. 4) were drawn using the *LinfK_scatterhist* function of the *fishboot* package (Schwamborn *et al.* 2018a), based on 1,000 successful bootstrap runs.

Uncertainty in length-based growth estimates was assessed through bootstrapped length-frequency analysis (BLFA). For this purpose, 95% CIs of VBGF parameters were calculated by using the *ELEFAN_GA_boot* function in the *fishboot* R package (Schwamborn *et al.* 2018a), which is a bootstrapped version of the *ELEFAN_GA* curve fitting function in *TropFishR* (Mildenberger *et al.* 2017). The function *ELEFAN_GA_boot* was used with 1,000 bootstrap runs applying the following precision-optimized parameters: MA = 7, seasonalised = TRUE, maxiter = 100, run = 40, addl.sqrt = FALSE, parallel = TRUE, low_par = NULL, popsize = 50, pmutation = 0.2, low_par = list ($L_\infty$ = 35 mm, K = 0.01, t_anchor = 0, C = 0, ts = 0), up_par = list ($L_\infty$ = 1050 mm, K = 1, t_anchor = 1, C = 1, ts = 1), following the recommendations given in Taylor & Mildenberger (2017) and in Schwamborn *et al.* (2018b). For each bootstrap analysis, duration per run (min. run$^{-1}$) was recorded, using a common PC with six-core CPU (AMD FX-6300, 3.5GHz). Bivariate confidence contour envelopes (Fig. 4) were drawn using the function *LinfK_scatterhist* in the *fishboot* package (Schwamborn *et al.* 2018a, 2018b, 2019), based on 1,000 successful bootstrap runs.

**Comparing the precision of methods**

To test for significant differences in the precision of methods (*e.g.,* BTA *vs* BLFA), a interquantile range test (Schwamborn 2019), *i.e.*, a non-parametric test for the comparison of inter-quantile ranges of bootstrap posteriors, was conducted, as described in Schwamborn et al. (2018b) and in Schwamborn (2019). The interquantile range test was used as implemented in the R function *interquant_r.test* within the *fishboot* package (Schwamborn *et al*. 2018a), at $p_{crit}$ = 0.05.

**Uncertainty in mortality estimates**

A two-step-bootstrap approach was used to assess uncertainty in mortality estimates. Step one was to bootstrap the *grotag* function to obtain posterior distributions for K and $L_\infty$ (see above). Step two was to apply these K and $L_\infty$ values to build LCCC plots and obtain a large number of Z values, i.e., a posterior distribution for Z. The posterior distribution for the Z central estimate considers only the uncertainty in the procedures leading to the LCCC, while the full, merged posterior distribution (posterior distributions for the Z central estimate, upper and lower 95%¨CI estimates for Z from the LCCC linear regression model) considers the uncertainty in both steps (pre-LCCC steps and LCCC linear regression model).

Z/K ratios were calculated based on posterior distributions of K from BTA and posteriors of Z from two-step-bootstrap with LCCC. These two posterior distributions were combined with a simple Monte Carlo approach, where a sample (n = 1) was taken randomly from each of the two posterior distributions, to calculate a set of unique Z/K ratios. This was done repeatedly (1,000 runs) and Z/K ratios were saved for posterior analysis. This posterior distribution of 1,000 Z/K values was used to calculate CIs for Z/K.

In this study, all 95% CIs were calculated based on the 0.025 and 0.975 quantiles of posterior distributions, using the *quantile* function in standard R (R Development Core Team 2019).

**Results**

**Length-frequency analysis**

A total of 1,078 individuals of *C. guanhumi* were caught and measured during the study period. Carapace width (CW) ranged from 20.9 mm to 70.0 mm, with a mean of 43.4 mm (standard deviation: 8.5 mm, median: 44.0 mm). Total weights varied between 4.0 g and 162.0 g, with a mean of 45.8 g (standard deviation: 25.3 g, median: 44.0 g). Sex ratio between 572 males (53%) and 506 females (47%) was significantly different from equality (sex ratio = 1.13: 1, $\chi^2$ = 3.93, p = 0.047). No significant differences in size or weight parameters were detected between sexes (permutation test, p > 0.05). Thus, all analyses were conducted with pooled (males and females) data. The vast majority (99.2 %) of all captured individuals were adults, with only 14 juveniles.

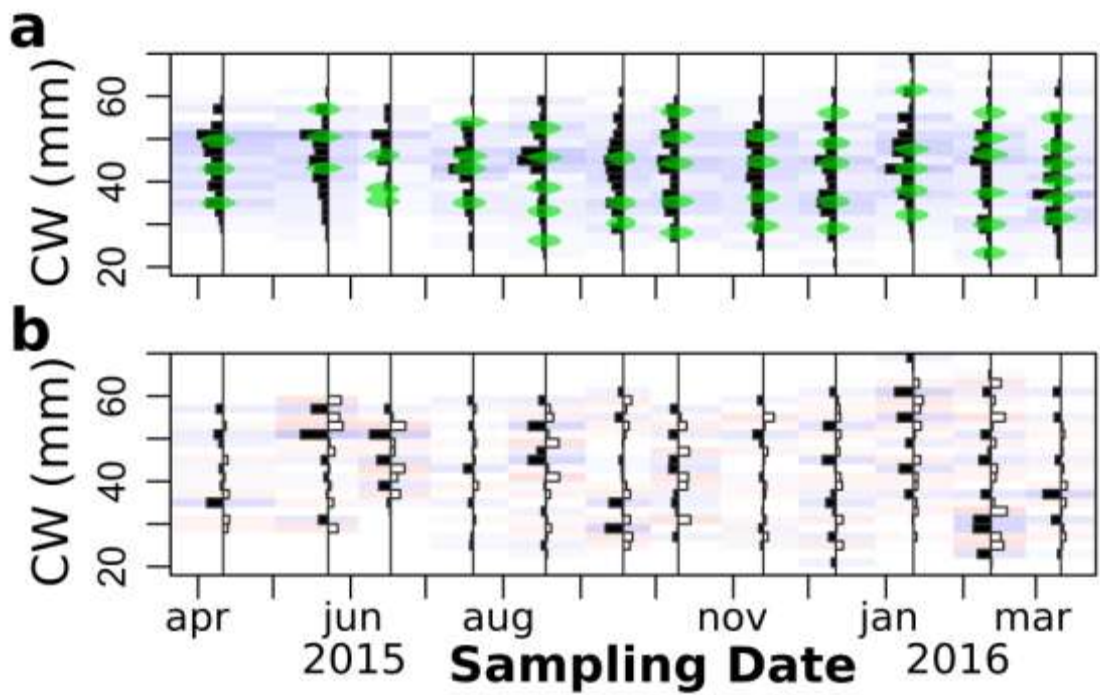

**Figure 1. Length-frequency distributions of 1,078 individuals of land crabs (*Cardisoma guanhumi*) caught in the Itamaracá mangroves, Brazil.** Above ("a"): raw length-frequency distributions (black bars) with cohorts detected by the Bhattacharya method (green dots). Below ("b"): restructured data with MA = 7.CW: carapace width.

Multi-modal distributions were observed in all months, with up to seven distinct modes per month (Fig. 1). There was no evident, unique modal progression to be observed in these data, whether using Bhattacharya,

Shepherd's method, ELEFAN I in FISAT II, or the *ELEFAN_GA* function in *TropFishR*. For instance, the Bhattacharya method produced a complex map of multiple modes (green dots in Fig. 1), which can be subjectively connected in numerous ways, offering many different potential growth curves. Shepherd's method and ELEFAN I also provided several possible solutions (growth curves), with nearly identical goodness of fit (Rn) values.

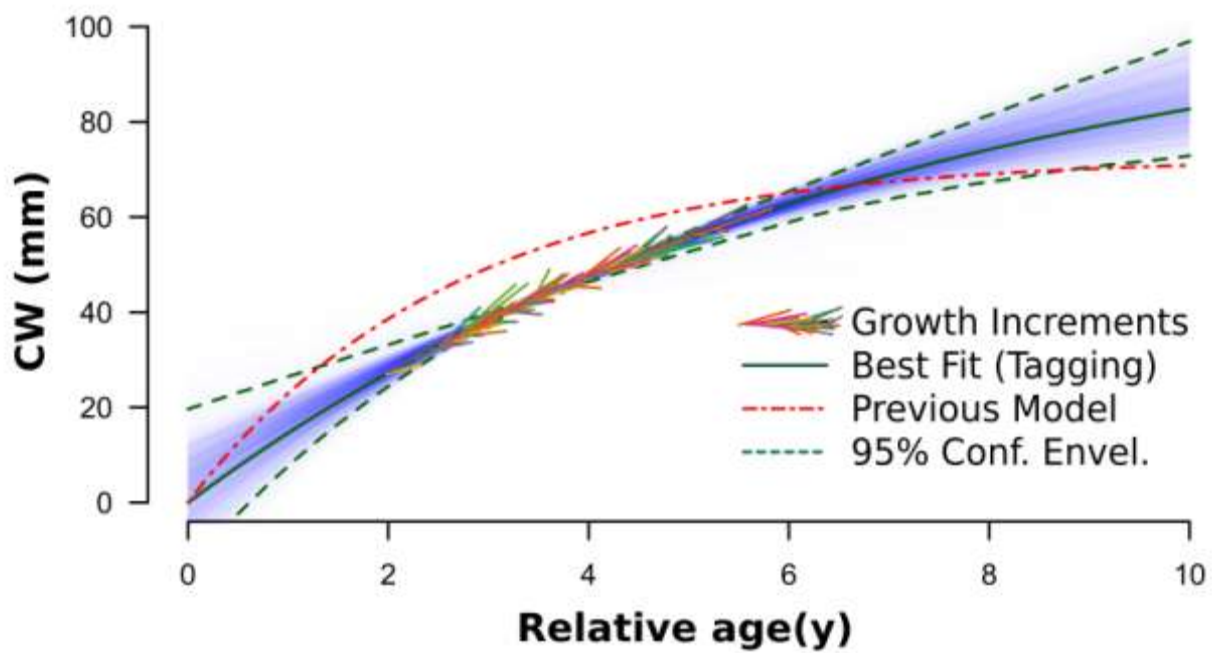

**Figure 2. Growth curves for land crabs (*Cardisoma guanhumi*) in the Itamaracá mangrove, Brazil, obtained by traditional length-frequency analysis (red) and with bootstrapped marc-recapture data (green line with gray envelope).** Continuous green line: VBGF curve with "best fit" parameters obtained by mark-recapture ($L_\infty$ = 108.03 mm, K = 0.145 $y^{-1}$), based on 130 size increments of tagged individuals. Blue dashed curves and grey area: 95% confidence envelope obtained by bootstrapping, based on marc-recapture data. Red dashed line: Previous VBGF curve, obtained with common length-based methods (Powell-Wetherall plot and K-scan, $L_\infty$= 72.5mm, K = 0.38 $y^{-1}$).

Standard optimization routines using a P-W plot and subsequent ELEFAN I with K-scan or Response Surface Analysis (RSA) did not produce a single optimum result, neither. When using the FISAT II software, the P-W method produced a Z/K ratio of 4.2 and a $L_\infty$ estimate of 75.9 mm (close to the $L_{max}$ value of 70 mm). When using *TropFishR*, $L_\infty$ estimates obtained from P-W plots varied from 69.3 to 75.9 mm (depending on subjective point selections). When using the P-W method in *TropFishR*, 95 % CIs for $L_\infty$ varied from 48.9-89.7 to 30.6-119.2 mm, and Z/K estimates varied from 3.1 to 4.8 (95% CIs for Z/K varied from ZK 3.1-3.2 to 4.7-4.9).

The K-Scan routine in FISAT II yielded three peaks of approximately equal height, with three possible "optimum" K values: 0.15, 0.21, 0.38 (all had Rn Scores of 0.20 to 0.21). When using the traditional RSA method (within FISAT II and *TropFishR*), the best VBGF growth model ($L_\infty$ = 72.5 mm; K = 0.38) had a Rn Score of 0.21, and was similar to one of the 'optimum' results obtained by K-Scan. The optimal $L_\infty$ estimate obtained with RSA within FISAT II ($L_\infty$ = 72.5 mm) was very close to $L_{max}$ (70 mm) and to the $L_\infty$ estimate obtained with Powell-Wetherall plot method (75.9 mm) within FISAT II.

The optimization routine *ELEFAN_GA* within *TropFishR*, that was used to fit VBGF curves within a nearly unconstrained $L_\infty$ and K search space, produced numerous possible results, too, depending on the initial seed values. The "best fit" varied strongly between subsequent optimization runs, with strongly varying "best fit" estimates, showing the tendency of this optimization method to become trapped in a different local maximum within the RSA space in each run, even when using very time-consuming and precise optimization settings (e.g., maxiter = 100). Best fit estimates obtained for $L_\infty$ varied between 67.4 and 123.9 mm (mean: 91.3 mm, median: 84.58 mm) and K values between 0.076 and 0.5 $y^{-1}$ (mean: 0.199 $y^{-1}$, median: 0.151 $y^{-1}$). Rn scores varied from 0.201 to 0.294. After 100 subsequent optimization runs (i.e., after 100 runs * 100 iterations = 10,000 iterations), the best non-seasonal VBGF fit parameters obtained with ELEFAN_GA were $L_\infty$ = 110.6 mm, K = 0.094 $y^{-1}$, t_anchor = 0.741 and Rn Score = 0.294, indicating very slow growth, much slower than from traditional methods, and very large $L_\infty$, much larger than from the P-W method and considerably larger than local $L_{max}$ (70 mm).

**Mark-recapture results**

Among the 291 individuals that were successfully tagged with PIT tags (153 males and 138 females), 95 individuals were recaptured at least once. Tagging yielded 130 useful growth increments (dL/dt), where 84 individual increments were obtained from males and 46 from females. No significant differences in growth increments were detected between sexes (permutation test, p > 0.05).

The number of increments (130) was larger than the number of recaptured individuals (95), since several individuals were recaptured more than once. The maximum number of times a single individual was recaptured was five. Sixty-six individuals (69%) were recaptured twice; twenty-two individuals (23%) were recaptured three times; six individuals (4.6 %) were recaptured four times, and one individual (1%) was recaptured five times.

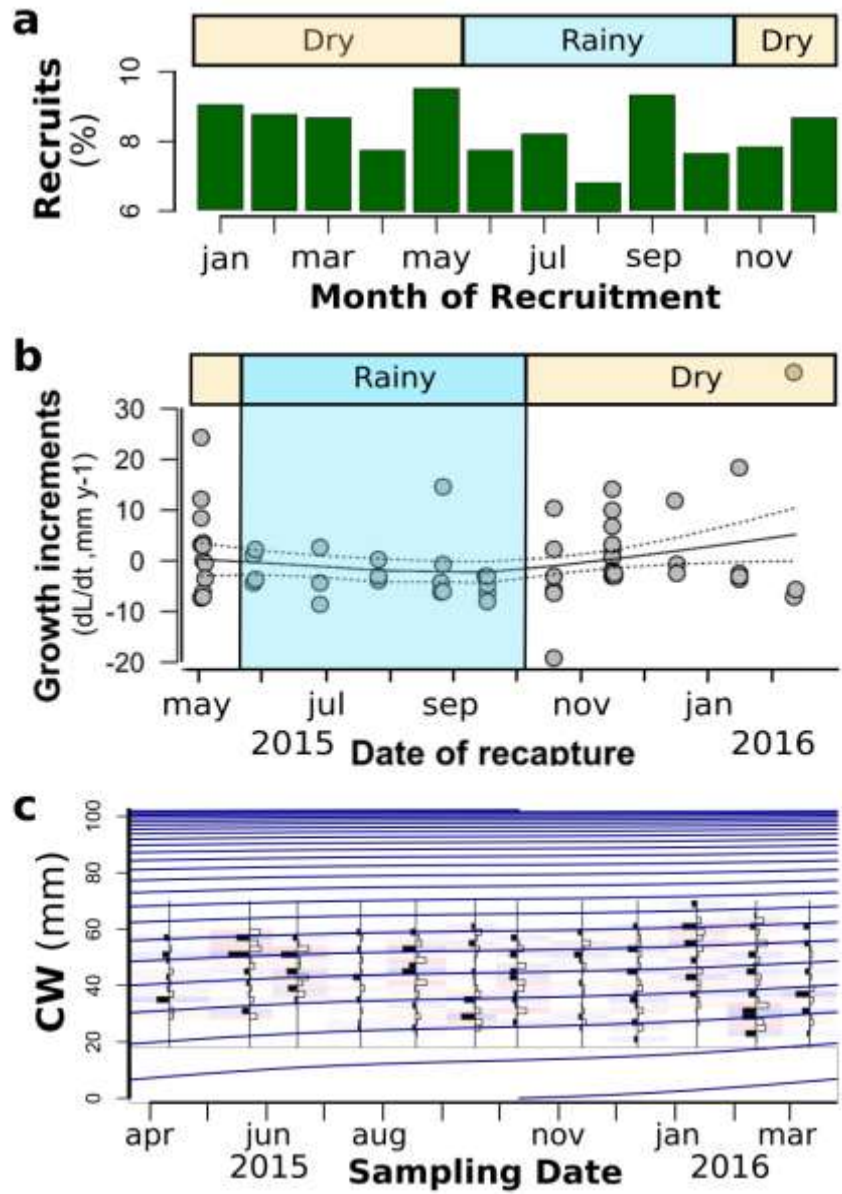

**Figure 3. Seasonality in recruitment and growth of land crabs (*Cardisoma guanhumi*) in the Itamaracá mangroves, Brazil**. a.) Seasonality of recruitment (%). b.) Seasonal variation of growth increments (dL/dt) obtained by marc-recapture. Lines: general additive model (GAM). c.) Example of a seasonally oscillating soVBGF growth curve ($L_\infty$ = 107.1 mm, K = 0.135 y-1, C = 0.53, t_anchor = 0.70, ts = 0.22, Rn = 0.28). Growth parameters were obtained by using the "best fit" parameters obtained by mark-recapture as seeds for final adjustments using ELEFAN_GA.

All 130 size increments could be combined into one single VBGF curve using the *grotag* function in *fishmethods* without any adjustments or exclusion of outliers (green line in Fig. 2). Conversely, the *growth_tagging* function within *TropFishR* did not converge towards useful results, producing negative numbers or results at the upper margin of any predefined search space.

The optimum growth parameters for all mark-recapture data (males and females) given by using *fishmethods* were K = 0.145 y$^{-1}$; L$_\infty$ = 108.03 mm. Growth parameters were very similar between sexes (L$_{\infty males}$ = 105 mm; L$_{\infty females}$ = 109 mm; K$_{males}$ = 0.16 y$^{-1}$; K$_{females}$ = 0.13 y$^{-1}$). The age of the individuals captured ranged from 1.5 years (20.9 mm) to 7.0 years (70.0 mm). L$_\infty$ obtained from mark-recapture (108 mm) was very similar to the best L$_\infty$ value obtained with ELEFAN_GA (110 mm) and considerably larger than values obtained from traditional LFA (P-W plot in FISAT II, L$_{\infty Powell-Wetherall}$ = 75.9 mm). Accordingly, the estimate of K from mark-recapture was much lower than from traditional LFA methods, leading to entirely different shapes of the growth curves obtained with these two methods (green line *vs* dashed red line in Fig. 2).

**Seasonal recruitment and growth**

Recruitment was continuous throughout the year, with lower recruitment during the rainy season (Fig. 3a). Seasonal variation in growth was well described with a non-linear GAM model (Fig. 3b), which was significantly different from zero (p = 0.022). Pairwise tests also showed that increments were significantly smaller in the rainy season than in the dry season (p = 0.006, n = 44 increments, Mann-Whitney test). Mean growth in the dry season was 5.1 mm y$^{-1}$ (st. dev.: 9.8 mm y$^{-1}$). Conversely, mean growth in the rainy season was negative, with -1.1 mm y$^{-1}$ (st.dev.: 2.9 mm y$^{-1}$).

**Fitting a seasonally oscillating growth curve**

When applying the seasonally oscillating soVBGF, the best fit also varied strongly between subsequent ELEFAN_GA runs, with strongly varying 'optimum' estimates, showing the tendency of this optimization method to become trapped in a different local maximum within the search space in each run (when using randomly varying seed values), even when applying very time-consuming optimization settings (e.g., maxiter = 100).

Best fit soVBGF parameter estimates with nearly unconstrained C (C varying from 0 to 1), using the search algorithm ELEFAN_GA varied strongly between runs, depending on the seed values used. "Best fit" L$_\infty$ estimates were generally much lower than Lmax (70 mm) and about half the optimal $L_\infty$ of non-seasonal LFA with ELEFAN_GA (110.6 mm). Most of the "best fit" $L_\infty$ estimates obtained for the soVBGF were below 55 mm, and 10 % of the "best fit" estimates were below 53.3 mm.

"Best fit" estimates for $L_\infty$ varied between 51.03 and 115.16 mm (mean: 67.98 mm, median: 55.11 mm), K varied between 0.074 and 0.92 y-1 (mean: $0.41y^{-1}$, median: 0.42 $y^{-1}$), and C between 0.13 and 0.96 (mean: 0.57, median: 0.45). Rn scores varied from 0.25 to 0.39.

After 100 subsequent optimization runs (i.e., after 100 runs * 100 iterations = 10,000 iterations), the optimum ('best of the best", with overall highest Rn score) soVBGF fit parameters $L_\infty$ and K were very similar to the optimum parameters for non-seasonal VBGF growth. Optimum soVBGF parameters obtained with ELEFAN_GA were $L_\infty$ = 115.16 mm, K = 0.088 $y^{-1}$, C = 0.60, ts = 0.575, t_anchor = 0.619, Φ' = 1.07 $\log_{10}(y^{-1})+\log_{10}(cm)$ and a very high Rn score (Rn = 0.389).

The growth curves obtained with LFA, applying a fixed (or narrowly constrained) seasonal growth amplitude C (derived from comparisons of mark-recapture increments between seasons), were very close to the to the optimum LFA results obtained with nearly unconstrained C. The combined use of several methods showed that an independent estimate of C, obtained from mark-recapture, could be successfully applied to LFA (see example in Fig. 3c). When applying a narrow search range for C, obtained from mark-recapture (C = 0.47 to 0.57), the soVBGF with ELEFAN_GA, with and all other parameters nearly unconstrained (i.e., searching within extremely wide search spaces, e.g. $L_\infty$ = 35 to 1050 mm; K = 0.01 to 1 $y^{-1}$; ts = 0 to 1), resulted in an optimum curve fit with a very large $L_\infty$ and very slow growth: with $L_\infty$ =119.3 mm; K= 0.084 $y^{-1}$; t_anchor = 0.564; C = 0.514; ts = 0.562; Φ' =1.08  1.07 $\log_{10}(y^{-1})+\log_{10}(cm)$ and a very high Rn score (0.383).

**Bootstrapped tagging analyses (BTA)**

Bootstrapped tagging analysis with *grotag_boot* (*fishboot* package) was extremely time-efficient, with less than three minutes for 1,000 runs, with or without considering seasonality in growth. Median values of posterior distributions of $L_\infty$ and K based on 1,000 runs were very close to the K and $L_\infty$ estimates obtained from the original increments with the simple, non-bootstrapped *grotag* function. Median $L_\infty$ after bootstrapping was very large, far above $L_{max}$, and similar to the previous optimum value with $L_\infty$ =118.1 mm, (95% CI for $L_\infty$: 80.8 to 362.7 mm) and median K was 0.121 $y^{-1}$, (95% CI for K: 0.024 to 0.26 $y^{-1}$), median Φ' was 1.24 $\log_{10}(cm+y^{-1})$, (95%CI: 1.16 to 1.54 $\log_{10}(cm+y^{-1})$).

Including seasonality into mark-recapture analyses significantly reduced the precision of growth parameter estimates. Seasonal growth models were significantly less precise than non-seasonal growth models for K ($p < 0.001$, interquantile range test) and for $\Phi$' ($p = 0.024$, interquantile range test). However, for $L_\infty$, CI widths were not significantly different between seasonal and non-seasonal growth models (interquantile range test). Also, 95% CIs for the seasonal amplitude parameter "C" always included C = 0, and the median solution was always C = 0 ($C_{median} = 0.0$), indicating that the non-seasonal VBGF is the best model describing the combined mark-recapture data sorted along relative age (Fig. 2). After conducting 8,000 bootstrap resampling runs, 3,841 successful runs (48%) were obtained with useful random combinations of growth increments, where a VBGF could be fitted with *grotag_boot*. Among these, 1,000 results were randomly selected and used for further analyses and plotting (e.g., Fig. 2).

**Bootstrapped length-frequency analyses (BLFA)**

Bootstrap resampling from LFDs (i.e., boostrapped length-frequency analysis, BLFA), by applying the novel *ELEFAN_GA_boot* function (*fishboot* package), produced useful results in all runs (n = 1,000). Length-frequency-based 95% CIs, obtained from *ELEFAN_GA_boot*, were conspicuously and significantly wider ($p < 0.0001$ for all parameters, interquantile range test) than those obtained from tagging (Fig. 4). Tagging was approximately 3 times more precise than BLFA for the estimation of the growth coefficient K, and 2.2 times more precise for $L_\infty$. In spite of its higher complexity, the seasonal growth model was not more precise ($p > 0.05$ for K, $L_\infty$ and $\Phi$', interquantile range test) than the simple non-seasonal VBGF. This is in opposition to mark-recapture results, where the non-seasonal growth model provided more precise results (i.e., narrower CIs).

For the optimum seasonal BLFA growth model, median amplitude C was 0.56 (very similar to estimates of C from non-bootsrapped LFA), indicating strong seasonal oscillations in growth. The 95% CI for C did not include C = zero (95% $CI_C$: 0.15 to 0.93), indicating the existence of seasonality in growth, based on LFDs plotted at precise sampling dates. Median $L_\infty$ for the bootstrapped soVBGF was 133 mm (95% $CI_{Linf}$: 50 to 673 mm) and median K was 0.10 $y^{-1}$ (95% $CI_K$: 0.01 to 0.73 $y^{-1}$), median $\Phi$' was 1.29 $log_{10}(cm+y^{-1})$, (95% $CI_{\Phi'}$: 0.93 to 1.93 $log_{10}(cm+y^{-1})$).

Due to the complex genetic algorithms used for optimization, BLFA was much slower than tagging-based analyses. BLFA with non-seasonal VBGF growth took approx. 5h per 1,000 runs, the full seasonal soVBGF models were even slower, with a duration of approx. 20h per 1,000 runs. Furthermore, computation time was also affected by the amplitude of the search space (i.e., wideness of search ranges for K, $L_\infty$ and C), and by the *ELEFAN_GA* precision-related parameters "maxiter", "run", and "pmutation", used for search optimization.

**Bivariate confidence contour envelopes for K and $L_\infty$**

Bivariate confidence envelopes for growth parameters ($L_\infty$ and K) always displayed a characteristic "banana shape", *i.e.*, the confidence contours were elongated, strongly bent and spread along a specific $\Phi'$ isopleth (Fig. 4). The confidence envelopes obtained from mark-recapture and BLFA both followed this general shape, but with marked differences. Confidence envelopes derived from mark-recapture *(grotag_boot)* were conspicuously narrower for K ("flat" banana) than those derived from BLFA (Fig. 4). Conversely, the shape of the confidence envelope derived from BLFA (Fig. 4b) was relatively short for $L_\infty$ and very tall for K, illustrating a much lower precision for K (3 times less precise than K obtained from tagging), but still within a conspicuous alignment along a specific $\Phi'$ isopleth (Fig. 4b).

**Mortality estimates and their confidence intervals**

Total mortality Z, as estimated by the common LCCC method, was 2.18 $y^{-1}$. This high mortality, combined with slow growth (e.g., K = 0.12) gives an extremely high Z/K ratio (Z/K = 18). This Z/K ratio of 18 is much higher (almost four-fold) than the Z/K ratio estimated by the widely used P-W method (e.g., $Z/K_{Powell\text{-}Wetherall}$ = 4.2). Uncertainty in Z, when estimated only by the traditional regression-derived CI, was relatively low, with a relatively narrow CI (95% $CI_Z$ = 1.7 – 3.1 $y^{-1}$). CI width from the common LCCC linear regression model was 3.1 - 1.7 = 1.4 $y^{-1}$. If the LCCC was a perfectly straight line, this CI width would be zero.

Each unique combination of growth parameters K and $L_\infty$, used as input for LCCC, produced a unique Z estimate. These Z posteriors (left histogram in Fig 5b) could be used to calculate 95% CIs for Z. This two-step bootstrap, considers uncertainty from the LCCC regression and uncertainty in original growth data (using posteriors from mark-recapture). This approach provided a wider 95% CI for Z (95% $CI_Z$ = 1.5 to 3.5 $y^{-1}$), than when considering only the uncertainty derived from the LCCC regression model.

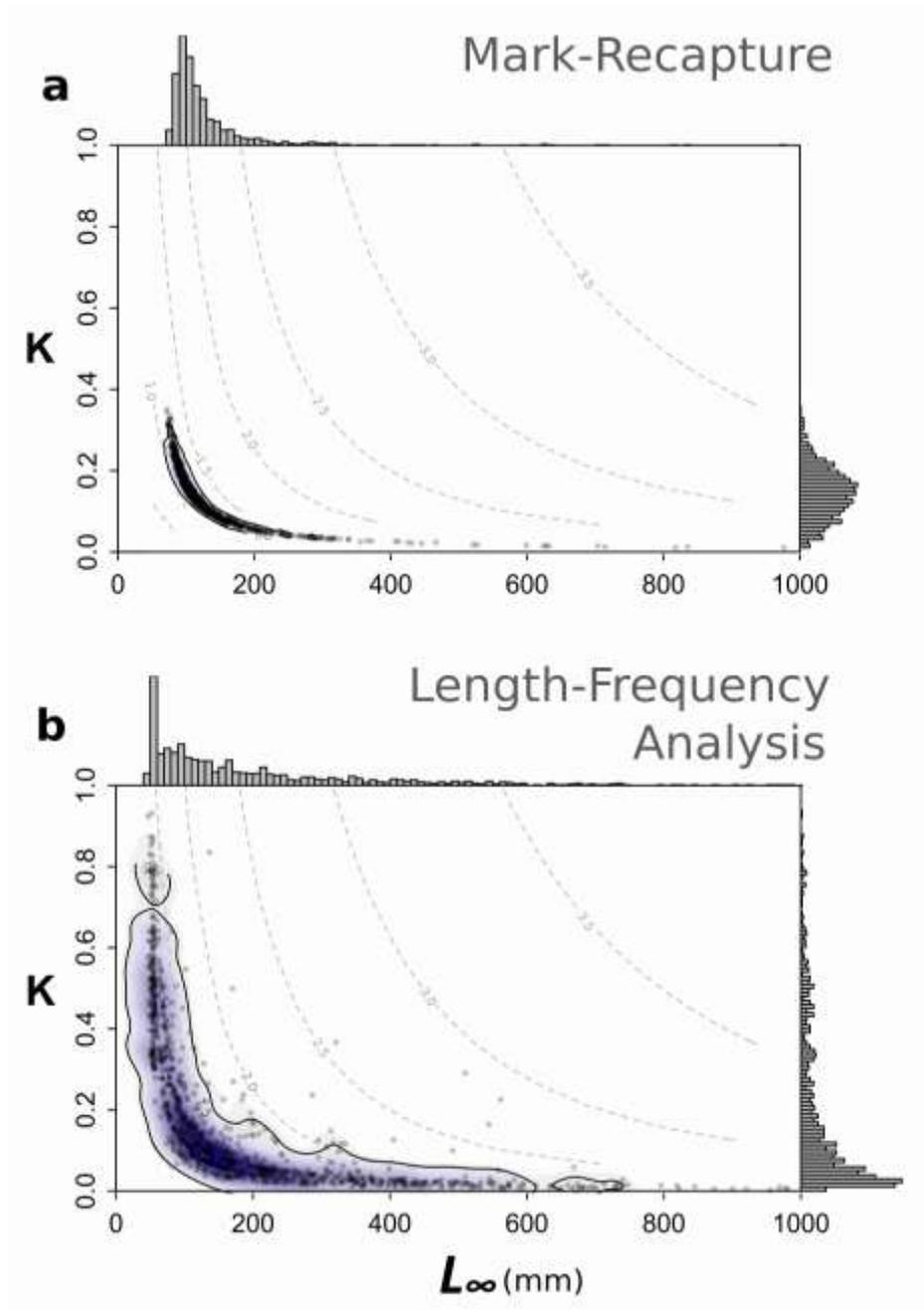

**Figure 4. Bivariate confidence envelopes for growth parameters K and L$_\infty$ for land crabs (*Cardisoma guanhumi*) in the Itamaracá mangroves, Brazil.** a.) Kimura plot based on tagging (130 growth increments). b.) Kimura plot based on length-frequency analysis with ELEFAN I (1,078 individuals). nboot = 1,000 bootstrap runs for both methods. Outer contour: 95% confidence envelope. Grey lines: Φ' isopleths.

CI width obtained from the LCCC with two-step bootstrap was 3.5 - 1.5 = 2 $y^{-1}$, considerably larger than CI width obtained from common LCCC only. This method considers both sources of uncertainty (LCCC linear regression and steps prior to LCCC). Thus, for this population of *C. guanhumi*, 70% (100 * (1.4 / 2) = 70%) of the total uncertainty in mortality was derived from the LCCC linear regression model (i.e., from irregularities in the shape of the catch curve), while the remaining 30% (100 * ((2-1.4) / 2) = 30%) of the total uncertainty were derived from prior steps (i.e., determination of K and $L_\infty$ by tagging).

The posterior distribution of Z/K ratios was calculated from randomly sampled pairs of bootstrap-derived posteriors of Z and K, thus considering uncertainty in Z and in K (95% $CI_K$ = 0.024 to 0.26 $y^{-1}$), and non-linear error propagation. The median of the posterior distribution was Z/K = 18, 95% $CI_{Z/K}$ = 7.5 to 98. The application of bootstrap-based methods revealed a considerably higher Z/K ratio and a much higher uncertainty in Z/K (i.e., a larger CI width) than the traditional Powell-Wetherall plot method (e.g., 95% $CI_{Z/K}$ = 3.1 to 4.9, using *TropFishR*), showing the importance of using robust, bootstrapped methods.

**Discussion**

In this study, a novel approach was applied to investigate a severely threatened land crab population, revealing slower growth, higher mortality and thus higher vulnerability to overharvesting than previously acknowledged. This is the first study to conduct bootstrapped tagging analysis (BTA) and to compare the precision and uncertainty of tagging and length-frequency analysis.

**A plea for bootstrapped, unconstrained optimization**

The first algorithms length-based analyses of body growth, such as ELEFAN I (Pauly & Gaschütz 1979; Pauly & David 1981) and MULTIFAN (Fournier *et al.* 1990) were created to improve earlier paper and pencil methods. There have been many recent advances since these pioneering efforts, such as the "*ELEFAN in R*" (Pauly & Greenberg 2013) and *TropFishR* packages (Mildenberger *et al.* 2017). Yet, the basic approach has remained the same since the very beginning: first, find the single "best" value for $L_\infty$ (using questionable methods, such as $L_{max}$ or P-W plot), then search for an optimum K value, keeping $L_\infty$ fixed.

Our results, based on two independent field methods (mark-recapture and length-based methods), indicate severe shortcomings and dangerous pitfalls in these highly popular approaches and methods. Estimates for $L_\infty$ from traditional methods ($L_{max}$ = 70 mm, $L_{\infty Powell\text{-}Wetherall}$ = 75.9 mm) would provide much smaller estimates of asymptotic size than any $L_\infty$ values obtained from BLFA and BTA. This confirms the conclusions of a recent simulation study (Schwamborn 2018), that is not possible to fix or constrain $L_\infty$ *a priori*, just by looking at the largest organism or by using a P-W plot. Fixing $L_\infty$ *a priori* by such dubious approaches has been widely used and was explicitly recommended by many authors (e.g., Gayanilo et al. 1997, Sparre & Venema 1998, Schmalenbach *et al.* 2011, Taylor & Mildenberger 2017), and may thus still be considered a current paradigm in fisheries science (Schwamborn 2018).

The use of traditional LFA approaches for this *C. guanhumi* population would lead to an underestimation of $L_\infty$ and subsequent drastic overestimation of K, as in previous studies in this region. An extremely biased result would also have been obtained when simply using $L_{max}$ (70 mm) to assess $L_\infty$, or, even worse, "the average of the ten largest" (Wetherall *et al.* 1987), which would give an even lower $L_\infty$ ($L_{average10max}$ = 62 mm) and erroneous subsequent calculations for growth and mortality. An underestimation $L_\infty$ and subsequent overestimation of K by traditional LFA methods is what would be expected for a population with Z/K ratio far above 2 ("Type B" population *sensu* Schwamborn 2018). Clearly, $L_\infty$ should not be estimated from local $L_{max}$ and not be fixed prior to model adjustment. Instead, $L_\infty$, K, and their uncertainties should be simultaneously assessed, explicitly considering their strong interrelationship.

Also, the present study shows that the uncertainty (i.e., the CI width) for $L_\infty$ may be much larger than previously thought, even when using an extremely accurate and reliable method for individual growth, such as mark-recapture with PIT tags. Furthermore, the Z/K ratio obtained from tagging (Z/K = 18, 95%$CI_{Z/K}$: 7.5 to 98) was several times higher than any estimate with the P-W method (Z/K estimates from 3.13 to 4.8). This confirms the severe bias in Z/K estimates obtained from the P-W method, with data of a real population, as recently suggested by the simulations of Hufnagl *et al.* (2012) and Schwamborn (2018). One especially serious pitfall of the P-W method is the absurdly narrow 95% CI for Z/K ratios (e.g. $Z/K_{powell\text{-}Wetherall}$ = 3.08-3.17, using *TropFishR*), that will lead to a dangerous overconfidence in these clearly erroneous results.

Another dreadful pitfall of popular LFA methods is their susceptibility to become trapped at a local maximum of the multimodal response surface (Schwamborn *et al.* 2018b). In the present study, repeated fit attempts with the modern curve fitting algorithm *ELEFAN_GA* (R package *TropFishR*) on the original land crab LFDs, applying different seed values (partial bootstrap *sensu* Schwamborn *et al.* 2018b), produced widely differing "best fit" VBGF parameters for each run, within a wide range of $L_\infty$ and K values. An incautious analyst, using always the same seed value, would most likely have obtained always the same, unique "optimum" result at a specific local maximum in repeated analyses, leading to erroneous overconfidence in apparently perfect and replicable results. The present study thus highlights the importance of conducting large numbers of repeated analyses for all possible combinations of parameters, using explicitly different seed values (maximum stochasticity), and bootstrapping within complex, infinite, multidimensional, multimodal search spaces.

**Comparison with previous studies on variability and uncertainty in body growth**

Until now, there was no simple routine available to measure uncertainty in population-level estimates of body growth rates, based on tagging. Only few tagging-based studies have provided detailed accounts of the variability in growth (Wang *et al.* 1995, Shakell *et al.* 1997, Zhang *et al.* 2009, Tang *et al.* 2014). Tang *et al.* (2014) used a complex hierarchical Bayesian model to analyze mark-recapture data from two species of freshwater mussels. They presented their K and $L_\infty$ posteriors as a biplot, based on Monte Carlo simulations, but without constructing bivariate confidence contour ellipses. Thus, the present study is the first to use a non-parametric bootstrap for mark-recapture analysis and to present biplots with well-defined confidence contour envelopes, derived from tagging data.

In this study, bivariate posteriors (K and $L_\infty$) were well aligned along Φ' isopleths, and accordingly, Φ' estimates in this study were very precise, for BTA and for BLFA. This is probably a general phenomenon: even under ideal circumstances, it is probably impossible to obtain precise and accurate estimates for K and $L_\infty$, from BLFA, due to their strong interdependence (higher K gives lower $L_\infty$ and vice-versa). At first sight, *a priori* fixing $L_\infty$ by dubious methods seems to "miraculously" solve this problem, but this will inevitably lead to a huge underestimation of the overall uncertainty and to severe bias in $L_\infty$, K, Z, and Z/K, as shown in Schwamborn (2018) and in the present study. Thus, a "slim" banana-shape, is probably the best possible situation for BLFD data, where estimates of K and $L_\infty$ are imprecise, but their interrelationship, described by Φ', can be very precise.

Several studies have presented posterior distributions in the form of K *vs* $L_\infty$ biplots, also called "Kimura plots", (Kimura 1980, Kingsford *et al.* 2019), in studies based on tagging (Tang *et al.* 2014, this study), otolith readings (Villegas-Ríos et al. 2013, Goldstein et al. 2016, Kingsford et al. 2019), LFD data (Schwamborn *et al.* 2018b, Herrón *et al.* 2018, this study), and simulations with perfect LFDs (Schwamborn *et al.* 2018b, Schwamborn *et al.* 2019).

Three types of Kimura plots can be distinguished regarding the uncertainty in K, $L_\infty$ and Φ': Type 1: perfect length-at-age data, with little variability in K and $L_\infty$ will produce a well defined, narrow ellipse in the Kimura plot, based on excellent otolith-reading data, where all length and age classes are well represented, from early juveniles to the oldest individuals (Kimura 1980, Goldstein *et al.* 2016, Kingsford *et al.* 2019), Type 2: "banana-shaped" confidence envelopes, where uncertainty for K and for $L_\infty$ is very high, but Φ' can be well assessed (Schwamborn *et al.* 2018b, Schwamborn *et al.* 2019, this study), and Type 3: "fried-egg-shaped" confidence envelopes derived from poorly structured LFDs (low pseudo-$R^2$ *sensu* Schwamborn *et al.* 2018b), as in Herrón *et al.* (2018), for Colombian Pacific fishes and for white clam (*Abra alba*) shell lengths in Schwamborn *et al.* (2018b).

The usefulness of Φ' has been intensively used to compare fish and other species since it was first proposed (Mathews & Samuel 1990, Murua *et al.* 2017). Here, we provide additional evidence for the usefulness for Φ', especially regarding the observation that Type 2 Kimura plots (with "banana-shaped" confidence envelopes) seem to be the best possible situation in LFA, and probably for some tagging-based analyses, too (this study). Precision and accuracy for K and $L_\infty$ in LFA is generally low, even for perfect synthetic LFD data (as in Schwamborn *et al.* 2019).

In the present study, the "banana" shape (Type 2 Kimura plot) obtained for *C. guanhumi* LFDs, in spite of the extremely wide 95% CIs, was conspicuously different from the amorphous "fried egg"-shaped biplots (Type 3) produced with the same methods, for LFDs of three fish species by Herrón *et al.* (2018) and for the white clam *Abra alba* by Schwamborn *et al.* (2018b). While in the present study, all biplots showed a unequivocal alignment to Φ' isopleths, in their data, there was no clear alignment of K and $L_\infty$ posteriors along a specific Φ' isoline (no "banana" shape), indicating a poor data structure (low signal/noise ratio and low "N")

Results from BLFA in the present study showed a high level of uncertainty for K and $L_\infty$ (huge CIs), and were not as perfectly aligned along a single Φ' isoline as for the perfectly shaped, synthetic LFDs presented in Schwamborn *et al.* (2019) and as in the well-aligned mark-recapture growth estimates presented here. In the present study, BTA was 2.2 to 3 times more precise than BLFA (narrower CIs for K and $L_\infty$), but even more remarkably, BTA also provided a better structure of the bivariate posteriors (posteriors were more narrowly aligned) resulting in narrower CIs for Φ'.

**Seasonality in growth**

In spite of the general assumption of seasonality in growth, including for populations in the tropics (Pauly & Gaschütz 1979, Pauly 1987, Laslett *et al.* 2004), the present study is the first to perform statistical tests for seasonality in growth within bootstrapped analyses of growth. All tests confirmed the existence of significant seasonality in growth in *C. guanhumi*, in spite of a very small seasonal thermal amplitude in air temperature, of only 3.2 °C, in the tropical coastal region of Itamaracá.

A relationship between seasonal growth amplitude C and seasonal thermal amplitude ΔT has first been suggested by Pauly & Gaschütz (1979). This relationship was further analyzed by Pauly (1987), where C = 0.11 *ΔT. In the case of *C. guanhumi* in the Itamaracá mangroves, the expected value for C, considering only the thermal amplitude, would thus be C = 0.11 * 3.2 = 0.35. However, all our estimates of C, whether by comparing mark-recapture growth increments between seasons, by LFA or by BLFA, resulted in considerably higher estimates for C (up to C = 0.6). It must be noted, however, that even Pauly (1987) suggested to use the estimates of C derived from ΔT as seed values for LFA only, not as conclusive values.

One reason for the large seasonal growth amplitude in *C. guanhumi* may be the observation that additionally to temperature, rainfall has a strong effect on the behavior and feeding of this species. These land crabs tend to be restrained inside their burrows during rainy periods, which obviously hampers their foraging activity. Thus, rather than being determined by temperature alone, seasonal feeding rhythms may be an important factor for these land crabs, such as for many other animals that have well-defined seasonal foraging rhythms. In many tropical coastal areas, seasonal factors other than temperature, such as rainfall and winds, may trigger changes in the environment and in the behavior of aquatic and land animals and thus in their growth rates. Another phenomenon, that certainly augments the seasonal growth oscillation in land crabs is the seasonal timing of the moulting cycle, with molts occurring predominantly in the dry season.

Considering the significant seasonality in growth, it may seem surprising, at first sight, that seasonal growth (soVBGF) was not an appropriate model for the description of mark-recapture increments, while for BLFA the seasonal model was the most precise. This is derived from the fact that individual growth curves are not perfectly synchronous regarding their recruitment and seasonal timing. While each individual most likely has a seasonally oscillating growth curve, the overlap in time produces an "average" non-seasonal growth curve, based on rearranged mark-recapture growth increments, where these rearranged increments loose their seasonal time stamp (Fig. 2). This explains why, for BTA, the non-seasonal VBGF was significantly more precise than the seasonal growth model. The issue of strong overlap due to non-synchronous recruitment and growth probably also applies to length-at-age analyses derived from otolith readings, where non-seasonal growth models are still widely used (Villegas-Ríos *et al.* 2013, Goldstein *et al.* 2016, Kingsford *et al.* 2019).

Conversely, for the analysis of time series of monthly LFDs, seasonality can be explicitly plotted and modeled while adjusting the growth curve on the progressing monthly cohorts, that maintain their exact sampling date stamp during growth curve adjustment. This explains why the soVBGF was more precise (smaller CIs) than the non-seasonal growth model for BLFA, in this study. Since the first applications, seasonal growth and mortality models have become widely accepted (Hufnagl *et al.* 2012). The approach to seasonality in the present study, with rigorous statistical tests, and combined BTA and BLFA, is a further incentive for the use of soVBGF models and a further evidence of the usefulness of LFD data for seasonality analyses in stock assessment and population studies, even when other, more precise and accurate data, are also available.

**Uncertainty in mortality estimates – consequences for stock assessment**

This study showed that the total mortality Z is strongly and log-linearly influenced by input Φ' values ($R^2$ = 0.98). The LCCC method seems to be robust to errors and variations in K and $L_\infty$, as long as lower K values lead to higher $L_\infty$, within slender Φ' isopleths, which is common for highly informative data (see above). Our simulations showed that Z was dependent on $L_\infty$, K, and Φ' when using independent data sets of K and $L_\infty$. Thus, when $L_\infty$ and K are not aligned along a Φ' isopleth (Type 3 bivariate datasets, e.g., poorly structured LFD data), any errors in estimates of $L_\infty$ (or in K) will affect the subsequent LCCC estimation of mortality directly (by the direct effect on the LCCC method). For Type 2 datasets, error in $L_\infty$ will be compensated by its effect on K, with constant Φ' and thus probably have little effect on the LCCC method. To ignore the uncertainty in underlying growth estimates and error propagation may lead to an underestimation in CI width for Z, which is a key parameter for any population studies or stock assessments.

In this study, with well-structured growth estimates ("Type 2" data pairs) obtained with BTA on PIT recapture growth increments, the 95% CI width of Z was very narrow (giving a precise estimate for Z), due to the compensatory nature of K and $L_\infty$ errors within LCCC (see above). Furthermore, only 30% of the overall uncertainty in Z for the *C. guanhumi* population in Itamaracá was derived from error propagation from growth estimates, the remaining 70% being due to the uncertainty that is intrinsic of the LCCC linear regression model, due to irregularities in the shape of the catch curve and the number of size classes used for regression. When using less precise growth estimates, such as those derived from BLFA, the percentage or error in Z derived from growth (i.e., from the initial estimates) in a complete evaluation is most likely much higher. This further highlights the need to acknowledge these sources of error, instead of only looking at the parametric error terms of the linear regression model, as done in most current analyses and popular stock assessment software tools.

In Itamaracá island, Z values for *C. guanhumi* were much higher and the CI width was much wider than reported for mangrove crabs *Ucides cordatus* in Pará state (Northern Brazil), investigated by Diele & Koch (2010). Extremely precise estimates for Z were obtained for *U. cordatus* in Pará, using LCCC: 0.69 ± 0.077 (95% CI) for males and 0.49 ± 0.086 (95% CI) for females (Diele & Koch 2010). These extremely narrow CIs are probably due to the "textbook" log-linearity of the *U. cordatus* catch curves and the fair numbers of large-sized mangrove crabs, providing many degrees of freedom for the linear regression model. Furthermore, the 95% CIs given by Diele & Koch (2010) did not consider the uncertainty derived from the underlying growth model and growth estimates, as done in this study.

Estimates of instantaneous natural mortality M (by definition, when there is no fishing, M = Z) obtained in the present study (Z = 2.2 $y^{-1}$, 95% CI = 1.7 to 4.5 $y^{-1}$), can be considered to be extremely high, since published estimates of M for fish and large crustaceans such as crabs and lobsters are generally below 1 $y^{-1}$, and values above M = 1 $y^{-1}$ are already considered to be high mortalities (Then *et al*. 2015, Costa *et al*. 2018). The 95% CI included values from high (Z = 1.7 $y^{-1}$) to extremely high (Z = 4.5 $y^{-1}$) mortality. Even with a very careful analysis and considering several sources of uncertainty, the 95% CI for mortality did not include a possibility of low mortalities (e.g., values of Z < 1), which indicates that this population is in a critical situation, even when considering the uncertainty in the methods and analyses used.

**Comparison with other *C. guanhumi* populations**

At first sight, the "dwarfed" size distribution of giant land crabs observed at Itamaracá island ($L_{max}$ = 70 mm) could seem to be a particularly dramatic case of a population that suffers from extremely high Z/K values. However, a quick survey of regional literature on this species shows that such small-sized *C. guanhumi* populations are very common in the tropical mangroves of northeastern Brazil. Several studies reported size distributions for northeastern Brazilian *C. guanhumi* populations that are similar or even smaller ($L_{max}$ = 62 mm, Botelho *et al*. 2001) than in the present study, sampled in mangroves in Pernambuco (Botelho *et al*. 2001), Paraíba (Takahashi 2008), Rio Grande Norte, (Silva 2013, Silva *et al*. 2014) and Ceará (Shinozaki-Mendes *et al*. 2013). Thus, the "dwarfed" *C. guanhumi* population analyzed in this study can be considered typical for this species in this region. Yet, in other regions, "giant" specimens do occur. In the considerably colder climate of subtropical southeastern Brazil, where *C. guanhumi* is not commercially exploited, large-sized animals have been reported, with $L_{max}$ values reaching 94 mm in São Paulo State (Gil 2009).

The present study agree with Cardona *et al*. (2019) who described a slow growth for *Cardisoma guanhumi* but differs from earlier attempts to assess body growth in *C. guanhumi* (Botelho *et al*. 2001), mainly in that all $L_\infty$ estimates obtained here are much larger than any individuals of this species ever reported from northeastern Brazil. This large $L_\infty$ estimate would seem suspicious in the eyes of any local analysts or ecosystem managers, who generally consider local or regional populations only. Looking only at previously reported $L_{max}$ and length-based $L_\infty$ values reported from this region with extremely overharvested populations, would lead to the erroneous conclusion that the large $L_\infty$ estimates (generally above 100 mm) obtained by mark-recapture in this study are unrealistic and thus incorrect. ( However, a wide-scale search proves that global $L_{max}$ for this species is far above local $L_{max}$ at Itamaracá Island (70 mm), especially when including pristine populations found in other regions. The $L_\infty$ estimates obtained in the present study are close to $L_{max}$ values reported from other regions, such as Florida ($L_{max}$ = 102 mm, Herreid 1963), Mexico ($L_{max}$ = 105 mm, Bozada and Chávez 1986, ) and Cuba ($L_{max}$ = 105 mm, Rivera 2005), where this species is not commercially caught. A quick search in SeaLifeBase (Palomares & Bailly 2011) yields an even larger $L_{max}$ value ($L_{max}$ = 150 mm), based on evidence from Florida (Hostetler *et al.* 2003), together with another very large $L_{max}$ value of 120 mm reported from Puerto Rico (Forsee & Albrecht 2012). Thus, to widen the search radius (in time and geographically) may help to open the perspective and to perceive the potential of any given species for its global maximum size. This may be helpful in defining plausible $L_\infty$ ranges in data-poor situations, especially for severely overfished and threatened populations, where current $L_{max}$ values are drastically reduced (Schwamborn 2018).

**Factors of mortality in land crabs**

Several factors may have contributed to the observed high mortalities in land crabs. Unsustainable harvesting by artisanal fishermen is a key aspect in this region, especially considering that harvesting of mangrove crabs (e.g., *Ucides cordatus*, *Goniopsis cruentata* and *C. guanhumi*) is the single most important source of income for the poorest fishermen in coastal communities in northeastern Brazil (Schwamborn & Santos 2009, Firmo *et al.* 2012). Yet, this aspect is possibly of lesser importance in closed areas (Silva 2013, this study), although occasional, illegal intrusions of fishermen in such areas may have occurred at some point in the years that preceded this study.

The most likely explanation for high Z is intensive predation by small mammals, such as stray dogs, stray cats, rats, crab-eating racoons (*Procyon cancrivorus*), opossum (*Didelphis* sp.), monkeys, and crab-eating foxes (*Cerdocyon thous*), all of which are very common in the study area (Moraes-Costa & Schwamborn 2018) and known to feed on mangrove crabs. Empty carapaces of *C. guanhumi* and other evidences of feeding by small mammals are abundant in the study area (Moraes-Costa & Schwamborn 2018). Predation mortality caused by small mammals, although not related to fisheries, can hardly be called "natural" mortality, since these predator populations are either introduced by humans (rats, stray pets, etc.) or positively affected by the removal of larger predators (large felines) by humans (Kuhnen *et al.* 2019). Another possible explanation for high crab mortalities can be diseases, such as the epidemic of yeast-like fungi of the family *Herpotrichiellacea* in 2004-2005, although such fungal pathogens have hitherto only been observed in crabs of the species *Ucides cordatus* (Vicente *et al.* 2012). Pesticides and other pollutants are probably also relevant in the context of crab mortality, especially considering that mangroves are commonly used as disposal sites for sewage, solid waste and numerous toxic chemicals (Yogui *et al.* 2018, Santos *et al.* 2019). The importance of harvesting by local fishermen for the population structure of *C. guanhumi* cannot be overstated, even in closed areas. This is especially evident when comparing the size structure of this species in northeastern Brazil and in regions where this species is not regularly harvested.

## Conclusions and Outlook

Most management decisions still assume the existence of unique, precise point estimates, even when there is a high variability in the underlying data, and consequently, there is considerable uncertainty in growth and mortality estimates obtained. Under such circumstances, doubtful approaches are often used for the fitting of growth curves (e.g., *a priori* fixing $L_\infty$). Instead, the present study highlights the importance of developing new management tools for data-poor stocks, that explicitly acknowledge the uncertainty within robust, replicable, multi-step, bootstrapped analyses.

## Declaration of Competing Interest

The authors declare that they have no known competing financial interests or personal relationships that could influence the content of this paper. Both authors are aware of the content of this work and agree to its publication.

## Acknowledgements

The authors thank the Center for Aquatic Mammals (CMA/CEPENE/ICMBio) for local support and for the authorization to perform this study in their mangrove patch, especially to F. Niemeyer, G. Sousa and F. Luna. The first author received a productivity fellowship from the Brazilian National Council for Scientific and Technological Development (CNPq), the second author received a M. Sc. fellowship (grant no. IBPG-0546-2.05/13) from the Foundation for the Support of Science and Technology of the State of Pernambuco (FACEPE). Many thanks to the National Institute of Science and Technology in Tropical Marine Environments - INCT AmbTropic (CNPq/FAPESB/CAPES) for support. The authors thank the Brazilian System of Authorization and Information in Biodiversity - SISBIO for the permit no. 43255-1. Special thanks to S. N. Leitão and all people who are part of UFPE's Zooplankton Laboratory for their help during fieldwork and to all who have somehow helped in this work. Many thanks to M. L. Palomares for encouraging the formation of a new working group on the subject of length-based methods and for her enthusiasm regarding this project. Many thanks to M. Taylor and T. Mildenberger for their dedication and competence in forming a new working group and in developing new bootstrap-based methods. Many thanks to Rainer Froese for important suggestions. Many thanks to D. Pauly for initiating this work many years ago and for numerous inspiring comments.

**Figure Legends**

**Figure 1. Map of the study area identifying sampling sectors A, B, C and D, at the upper fringe of the CMA mangrove at Itamaracá Island, Pernambuco, State, Brazil**

**Figure 1. Length-frequency distributions of 1,078 individuals of land crabs (*C. guanhumi*) caught in the Itamaracá mangroves, Brazil.** Above: raw length-frequency distributions (black bars) with cohorts detected by the Bhattacharya method (green dots). below: restructured data with MA = 7.CW: carapace width.

**Figure 2. Growth curves for land crabs (*C. guanhumi*) in the Itamaracá mangrove, Brazil, obtained by traditional length-frequency analysis (red) and with bootstrapped marc-recapture data (green line with gray envelope).** Continuous green line: VBGF curve with "best fit" parameters obtained by mark-recapture ($L_\infty$ = 108.03 mm, K = 0.145 $y^{-1}$), based on 130 size increments of tagged individuals. Blue dashed curves and grey area: 95% confidence envelope obtained by bootstrapping, based on marc-recapture data. Red dashed line: Previous VBGF curve, obtained with common length-based methods (Powell-Wetherall plot and K-scan, $L_\infty$= 72.5mm, K = 0.38 $y^{-1}$).

**Figure 3. Seasonality in recruitment and growth of land crabs (*C. guanhumi*) in the Itamaracá mangroves, Brazil**. a.) Seasonality of recruitment (%). b.) Seasonal variation of growth increments (dL/dt) obtained by marc-recapture. Lines: general additive model (GAM). c.) Example of a seasonally oscillating soVBGF growth curve ($L_\infty$ = 107.1 mm, K = 0.135 y-1, C = 0.53, t_anchor = 0.70, ts = 0.22, Rn = 0.28). Growth parameters were obtained by using the "best fit" parameters obtained by mark-recapture as seeds for final adjustments using ELEFAN_GA.

**Figure 4. Bivariate confidence envelopes for growth parameters K and $L_\infty$ for land crabs (*C. guanhumi*) in the Itamaracá mangroves, Brazil.** a.) Kimura plot based on tagging (130 growth increments). b.) Kimura plot based on length-frequency analysis with ELEFAN I (1,078 individuals). nboot = 1,000 bootstrap runs for both methods. Outer contour: 95% confidence envelope. Grey lines: Φ' isopleths.